\documentclass[11pt,dvips]{article}
cm \voffset = -2truecm

\usepackage{graphicx}
\usepackage{amssymb}
\usepackage{latexsym}
\usepackage{color}

\begin{document}
\begin{titlepage}
\setcounter{page}{1}
\renewcommand{\thefootnote}{\fnsymbol{footnote}}

\vspace{5mm}
\begin{center}

 {\Large \bf Quantum correlations dynamics of
 quasi-Bell cat states  }



\vspace{1.5cm}

{\bf M. Daoud}$^{a}${\footnote { email: {\sf
m$_{-}$daoud@hotmail.com}}}, {\bf R. Ahl Laamara}$^{b,c}${\footnote
{ email: {\sf ahllaamara@gmail.com}}} and {\bf R. Essaber}
$^b${\footnote { email: {\sf essaber.rim@gmail.com}}}

\vspace{0.5cm}
$^{a}${\it Department of Physics, Faculty of Sciences, University Ibnou Zohr,\\
 Agadir,
Morocco}\\[1em]

$^{b}${\it LPHE-Modeling and Simulation, Faculty  of Sciences,
University
Mohammed V,\\ Rabat, Morocco}\\[1em]

$^{c}${\it Centre of Physics and Mathematics,
CPM, CNESTEN,\\ Rabat, Morocco}\\[1em]

\vspace{3cm}

\begin{abstract}

A model of  dynamics of quantum correlations of two modes quasi-Bell
cat states, based on Glauber coherent states, is considered.
 The  analytic expressions of  pairwise entanglement of formation, quantum discord and its
geometrized variant are explicitly derived. We analyze the
distribution of quantum correlations between the two modes  and the
environment. We show that, in contrast with squared concurrence,
entanglement of formation, quantum discord and geometric quantum
discord do not follow the property of monogamy except in some
particular situations that we discuss.

\end{abstract}
\end{center}
\end{titlepage}

\newpage

\section{Introduction}

The idea to encode  information in multi-photon coherent states
constitutes a promising tool in quantum information. Indeed, the
coherent states superpositions have been employed as resource to
implement many quantum tasks including quantum teleportation
\cite{Enk,JKL01},
 quantum computation \cite{JK,Ralph,Ralph2},
entanglement purification \cite{JKpuri} and errors correction
\cite{Glancy}. These potential applications explain the special
attention paid, during the last years,  to the identification,
characterization and quantification of  quantum correlations in
bipartite coherent states systems (see for instance the papers
\cite{Sanders3,Sanders,Wang} and references therein). The bipartite
treatment was extended to superpositions of multimode coherent
states \cite{Jex,Zheng,Wang1,daoud1,daoud2} which exhibit
multipartite entanglement as for instance in {\rm GHZ}
(Greenberger-Horne-Zeilinger), {\rm W} (Werner) states
\cite{Jeong1,Li} and entangled coherent state versions of cluster
states \cite{Munhoz,Wang-WF,Becerra}. To quantify quantum
correlations beyond entanglement in coherent states systems,
measures such as bipartite quantum discord
\cite{Ollivier-PRL88-2001,Vedral-et-al} and its geometric variant
\cite{Dakic2010} were used. Explicit results were derived for
quantum discord \cite{Luo,Ali,Shi1,Girolami,Shi2,Rachid1,Rachid2}
and geometric quantum discord \cite{Adesso1,Giorda,Yin,Rachid3} for
some special sets of coherent states.\\
In other hand, decoherence is a crucial process to understand the
emergence of classicality in quantum systems. It describes the
inevitable degradation of quantum correlations due to experimental
and environmental noise. Various decoherence models were
investigated and in particular  the phenomenon of entanglement
sudden death  was considered in a number of distinct contexts (see
for instance \cite{YuEberly-2007} and reference therein). For
optical
 qubits based on coherent states,  the influence of the
environment,  is mainly due to energy loss or photon
 absorption. The photon loss or equivalently amplitude damping in a noisy
 environment can be modeled by assuming that some of field energy and
 information is lost after transmission through a beam splitter
 \cite{Rachid1,Wickert}. Another important issue in analyzing the
 decoherence process concerns the distribution of quantum
 correlations between the bipartite coherent states
 and the environment.  Accordingly, the study of the distribution of quantum correlations
in a quantum system among its different parts constitutes an
important issue. In fact, the free shareability of classical
correlations  is no longer valid in the quantum case and the
distribution of quantum correlations  obeys to severe restrictions.
These restrictions are known in the literature as monogamy
constraints. The concept of monogamy of entanglement for qubits was
first proved by Coffman, Kundo and Wootters in 2001 \cite{Coffman}
and since then it was extended to other measures of quantum
correlations \cite{Giorgi,Prabhu,Allegra,Ren,Bruss}. For a
tripartite system $ABE$, the monogamy relation can be presented as
follows. Let $Q_{A\vert B}$ (reps. $Q_{A\vert E}$ ) denote the
shared correlation $Q$ between $A$ and $B$ (resp. $A$ and $E$)  and
$Q_{A\vert BE}$ the correlation shared between $A$ and the composite
subsystem $BE$ comprising $B$ and $E$. The quantum correlation
measured by $Q$ is monogamous if
\begin{equation}\label{mono-relation}
Q_{A\vert BE} \geq Q_{A\vert B} + Q_{A\vert E}.
\end{equation}
In this paper, the focus will be maintained strictly on the
evolution quantum correlations  present in two modes quasi-Bell cat
states based on Glauber coherent states. We  study the monogamy
relation to understand the distribution of quantum correlations
 between the two modes of quasi-Bell cat states and the environment.
To approach this question, we use the bipartite measures:
entanglement of formation, quantum discord or geometric quantum
discord. This approach has the advantage relying upon bipartite
measures of entanglement of formation  and quantum discord that are
physically motivated and analytically computable.

This paper is organized as follows. In section 2, we introduce two
modes quasi-Bell cat states based on Glauber coherent states. We
discuss their evolution under amplitude damping modeled by the
action of a beam splitter.  We give the density matrices describing
the evolution of the two modes as well as ones describing each mode
coupled to the environment. In section 3, we explicitly derive the
entanglement of formation for each bipartite subsystem. We also
consider the distribution of entanglement between the system and the
environment. The explicit expressions of pairwise quantum discord
are  derived in section 4. We also consider the monogamy relation of
this measure which goes beyond entanglement of formation. Similar
analysis are presented in the section 5 when bipartite correlations
are measured  by means of the geometric discord. Concluding remarks
close this paper.

\section{Evolution of quasi-Bell states under amplitude damping }

\subsection{ Quasi-Bell states}
Usually the standard Bell states are constructed as balanced
superpositions of orthogonal states. Here, we consider
superpositions involving non orthogonal states. In particular, we
consider quasi-Bell states based on Glauber coherent states
\begin{equation}\label{glauber-cs}
|\alpha\rangle = e^{-\frac{|\alpha|^2}{2}} \sum_{n=0}^{\infty}
\frac{\alpha^n}{\sqrt{n!}}|n\rangle
\end{equation}
where $\vert n \rangle$ is a Fock state and $\alpha$ is the complex
amplitude of the coherent state. The four quasi-Bell cat states are
defined as
\begin{equation}\label{bell-cs}
\vert \alpha , \pm \alpha ; m \rangle = {\cal
N}_m^{-\frac{1}{2}}(\vert \alpha, \pm\alpha \rangle + e^{im\pi}
\vert - \alpha, \mp \alpha \rangle)
\end{equation}
where the normalization factor ${\cal N}_m$ is
$$ {\cal N}_m = ( 2 + 2 e^{-4\vert \alpha \vert^2}\cos m \pi  )$$
and the integer  $m = 0, 1 ~{\rm (mod~2)}$. Notice that   the two
modes quasi-Bell states can be converted in a state describing two
logical qubits. This can be realized by means of  even and odd
coherent states
$$ \vert \pm \rangle = {\cal N}_{\pm} (\vert \alpha \rangle \pm \vert -\alpha \rangle)$$
representing  the two superpositions of Glauber states of same
amplitude and opposite phase and ${\cal N}_{\pm}$ being the
normalization factors. The vectors $\vert + \rangle$ (even cat
state) and $\vert - \rangle$ (odd cat state) form an orthogonal
basis of two dimensional Hilbert space and then can be viewed as two
logical qubits. Furthermore, the even and odd Glauber coherent
states constitute an important resource in implementing
experimentally non orthogonal quasi-Bell states. For instance the
state $\vert \alpha , \alpha ; 0 \rangle$ can be produced by sending
a coherent cat state of the form
$|\sqrt{2}\alpha\rangle+|-\sqrt{2}\alpha\rangle$ and the vacuum into
the two input ports of a 50/50 beam splitter. Clearly, the
generation of quasi-Bell states requires a source of coherent cat
states. Some experimental achievements in this sense were obtained
recently (see for instance \cite{Yurke,Ralph03,Song,Lund} and
references therein). It is interesting to note that the quasi-Bell
states could be successfully employed for quantum teleportation and
many others quantum information processing \cite{Jeong2}. From this
perspective
 quantum optical states, including the quasi-Bell coherent states, are expected to be useful
 in the context of quantum information science, especially for
 communications using qubits over long distance. This is mainly
 motivated by the fact that  coherent states are more robust
 against photon absorption  (see
 \cite{Glancy}) and subsequently presents an advantageous alternative to reduce the
 decoherence effects.

\subsection{Photon loss mechanism of quasi-Bell cat states }

Coherent fields, traveling through a long optical channel, interact
inevitably with the environment. The coupling field-environment
changes the quantum correlations and consequently causes the
decoherence of the system. To characterize the environmental
 effects, we suppose that absorption of the transmitted
photons is the dominant source of the decoherence mechanism. The
description of the photon loss mechanism, also termed amplitude
damping, can be modeled by the action  of a beam splitter. In other
words, we assume that some of the coherent field is lost in transit
via a beam splitter. The coherent states enters one port of the beam
splitter and the vacuum $\vert 0 \rangle_E$, representing the
environment, enters the second port. After transmission some
information encoded in the coherent states is transferred and the
remaining amount of information is lost to the noisy channel. Let us
consider the system $AB$ (the two mode quasi-Bell states) and its
environment $E$ in initial state
$$\rho _{ABE}(0) = \rho _{AB}(0)\otimes \rho_E(0)$$
where
$$\rho _{AB}(0) = \vert \alpha , \pm \alpha ; m \rangle\langle \alpha , \pm \alpha ; m \vert
\qquad \rho _{E}(0) = \vert 0\rangle_E {_E}\langle 0 \vert.$$ The
dynamics of the whole closed system is unitary, i.e.,
$$\rho _{ABE} = {\cal U} ~\rho _{ABE}(0)~ {\cal U}^{\dagger}.$$
Two cases can be distinguished: the case that the two qubits only
interact with their local environments and  the case in which only
one qubit is affected by a local environment. The first case (reps.
the second) is called two-qubits (resp. one qubit) local amplitude
damping channel. Here,  we shall consider the situation where
 only the second mode of quasi-Bell cat states interact with the
 environment. In this scheme, we write the unitary operator describing the dynamical evolution
 of the whole system as
$$ {\cal U} =  \mathbb{I}\otimes{\cal B}(\theta)$$
where $\mathbb{I}$ is the identity and the beam splitter operator,
describing the interaction between the subsystem $B$ and the
environment $E$, is given
\begin{equation}\label{B-oper}
{\cal B}(\theta) = \exp\left[\frac{\theta}{2}  \left(a^-_B a^+_E -
a^+_B a^-_E\right)\right].
\end{equation}
 The objects $a^+_L$ and $a^-_L$ $( L = B , E)$ are the usual harmonic
oscillator ladder operators acting on the Fock modes of the
subsystems $B$ and $E$. The reflection and transmission coefficients
are
\begin{equation}
 t = \cos\frac{\theta}{2}~, \qquad  r = \sin\frac{\theta}{2}
\end{equation}
in terms of  the angle $\theta$ of the equation (\ref{B-oper}). The
beam splitter transmissivity describes the decoherence behavior of
the transmitted states. It can be related to the exponential energy
loss of an optical fiber used in the transmission process as $t = e
^{-\lambda L}$ where $\lambda$  is  a parameter characterizing  the
energy loss of the fiber over a distance $L$. The dynamical
evolution of the initial state under the action of the beam splitter
writes as
$$
\vert Q \rangle_{ABE} = (\mathbb{I}\otimes {\cal B}(\theta))\vert
\alpha , \pm \alpha ; m \rangle_{AB} \otimes \vert 0 \rangle _{E}.
$$
 It is simple to check that
\begin{eqnarray}
\vert Q \rangle_{ABE} = \frac{1}{\sqrt{{\cal N}_m }}(\vert \alpha
,\pm \alpha t, \pm \alpha r \rangle +e^{im\pi }\vert -\alpha ,\mp
\alpha t, \mp \alpha r \rangle ).
\end{eqnarray}
The whole system is then represented by the density matrix
\begin{eqnarray}
\rho _{ABE}= \vert Q \rangle _{ABE}\langle Q \vert = \frac{1}{{\cal
N}_m }\bigg(\vert \alpha ,\pm \alpha t, \pm \alpha r \rangle \langle
\alpha ,\pm \alpha t, \pm \alpha r \vert +e^{im\pi } \vert -\alpha
,\mp \alpha t, \mp \alpha r \rangle \langle \alpha ,\pm \alpha t,\pm
\alpha r \vert \nonumber \\+ \vert -\alpha ,\mp \alpha t, \mp \alpha
r \rangle \langle -\alpha ,\mp \alpha t,\mp \alpha r \vert +
e^{-im\pi } \vert \alpha ,\pm \alpha t,\pm \alpha r \rangle \langle
-\alpha ,\mp \alpha t,\mp \alpha r \vert \bigg)
\end{eqnarray}
It is important to emphasize that the environment is constituted by
the universe minus the subsystems $A$ and $B$. The density matrix
$\rho _{ABE}$ is  pure. As we shall be concerned with the
distribution of quantum correlations in this pure tripartite system,
we  denote by $\rho _{AB}$ the reduced states for the subsystems $A$
and $B$ and analogously for $\rho _{AE}$ and $\rho _{AE}$. After
tracing out all the modes of the environment, one gets
\begin{eqnarray}\label{rhoAB}
\rho_{AB} = \frac{{\cal N}_m (t)}{{\cal N}_m } \bigg[ \frac
{1}{2}(1+c_r)|\alpha , \pm \alpha t ; m \rangle\langle \alpha , \pm
\alpha t ; m| + \frac {1}{2}(1-c_r) Z |\alpha , \pm \alpha t ;
m\rangle\langle \alpha , \pm \alpha t ; m| Z\bigg]
\end{eqnarray}
where the $r$-dependant quantity $c_r$ is
$$c_r = e^{-2 r^2 \vert \alpha \vert^2}$$
 and the  states $\vert \alpha , \pm \alpha t ; m\rangle$ are given
 by
$$\vert \alpha , \pm \alpha t ; m\rangle =
{\cal N}_m (t))^{-\frac{1}{2}}(\vert \alpha, \pm \alpha t \rangle +
e^{im\pi} \vert - \alpha, \mp \alpha t \rangle)$$ with
$$ {\cal N}_m (t) = \bigg( 2 + 2 e^{-2(1+t^2)\vert \alpha \vert^2}\cos (m \pi ) \bigg).$$
The third Pauli operator $Z$ in (\ref{rhoAB}) is defined by
$$Z \vert \alpha , \pm \alpha t ; m\rangle =
{\cal N}_m (t)^{-\frac{1}{2}}(\vert \alpha, \pm\alpha t \rangle -
e^{im\pi} \vert - \alpha, \mp \alpha t \rangle).
$$
Similarly, tracing out the degrees of freedom of the subsystem $B$,
one finds
\begin{eqnarray}\label{rhoAE}
\rho_{AE} = \frac{{\cal N}_m (r)}{{\cal N}_m } \bigg[ \frac
{1}{2}(1+c_t)|\alpha , \pm \alpha r ; m \rangle\langle \alpha , \pm
\alpha r ; m| + \frac {1}{2}(1-c_t) Z |\alpha , \pm \alpha r ;
m\rangle\langle \alpha , \pm \alpha r ; m| Z\bigg]
\end{eqnarray}
where ${\cal N}_m (r)$, $c_t$ and the operation $Z$ are defined as
above modulo the obvious substitution $r \longleftrightarrow t$.
It is also simply  verified that the reduced state $ \rho_{BE} =
{\rm Tr_A}~ \rho _{ABE}$ is given by
\begin{eqnarray}\label{rhoBE}
\rho_{BE} =\frac{{\cal N}_m(0)}{{\cal N}_m}\bigg[\frac
{1}{2}(1+c_1)|\alpha t, \pm \alpha r ; m\rangle\langle \alpha t ,
\pm \alpha r ; m| + \frac {1}{2}(1-c_1) Z |\alpha t , \pm \alpha r ;
m\rangle\langle \alpha t , \pm \alpha r ; m| Z\bigg].
\end{eqnarray}
where ${\cal N}_m(0)= {\cal N}_m(t=0)$ and
$$|\alpha t, \pm \alpha r ; m\rangle = {\cal
N}_m^{-\frac{1}{2}}(\vert \alpha t , \pm\alpha r \rangle + e^{im\pi}
\vert - \alpha t, \mp \alpha r \rangle)$$ and $c_1 = e^{-2 \vert
\alpha \vert^2}$. \\
Having expressed the reduced density matrices of the different
subcomponents of the system quasi-Bell cat states coupled to its
surroundings, we shall consider, in the following sections, the
explicit evaluation of pairwise quantum correlations measured by
entanglement of formation, usual quantum discord and  geometric
quantum discord. A special attention will be paid to the monogamy
relation of each of these measures.

\section{Entanglement of formation}

To begin our task, we first derive the explicit expressions of
entanglement of formation measuring the bipartite correlations
present in the states $\rho_{AB}$, $\rho_{AE}$ and $\rho_{A\vert
BE}$ to discuss the entanglement monogamy measured by the
concurrence and entanglement of formation \cite{conc}. For this, we
shall map each of these bipartite subsystems in a pair of two
logical qubits. This mapping is based on the fact, as mentioned
above, that Shr\"odinger cat (even and odd) coherent states can be
identified with two orthogonal qubits.

\subsection{ Concurrence and entanglement of formation}
For the state $\rho_{AB}$, a qubit mapping  can be introduced as
follows. For the first mode $A$, we introduce a two dimensional
basis spanned by the vectors $\vert u_{\alpha} \rangle$ and $\vert
v_{\alpha} \rangle$ defined by
\begin{equation}\label{qubitA}
\vert \alpha \rangle = a_{\alpha} \vert u_{\alpha} \rangle +
b_{\alpha}\vert v_{\alpha} \rangle \qquad \vert -\alpha \rangle =
a_{\alpha} \vert u_{\alpha} \rangle - b_{\alpha}\vert v_{\alpha}
\rangle
\end{equation}
where
$$\vert a_{\alpha} \vert^2 + \vert b_{\alpha} \vert^2 = 1\qquad
\vert a_{\alpha} \vert^2 - \vert b_{\alpha} \vert^2 = \langle
-\alpha \vert \alpha\rangle.$$ To simplify our purpose, we take
$a_{\alpha}$ and $b_{\alpha}$ reals such as
$$a_{\alpha} = \frac{\sqrt{1 + p}}{\sqrt{2}} \quad b_{\alpha} = \frac{\sqrt{1 - p}}{\sqrt{2}} \qquad {\rm with}
 \quad p = \langle -\alpha \vert \alpha\rangle = e^{-2\vert \alpha \vert^2}.$$
Similarly, for the second mode $B$,  a two-dimensional basis
generated by the vectors $\vert u_{\alpha t} \rangle$ and $\vert
v_{\alpha t} \rangle$  is defined as
\begin{equation}
\vert \alpha t \rangle = a_{\alpha t} \vert u_{\alpha t} \rangle +
b_{\alpha t}\vert v_{\alpha t} \rangle \qquad \vert -\alpha t
\rangle = a_{\alpha t} \vert u_{\alpha t} \rangle - b_{\alpha
t}\vert v_{\alpha t} \rangle
\end{equation}
where
$$a_{\alpha t} = \frac{\sqrt{1 + p^{t^2}}}{\sqrt{2}} \quad b_{\alpha t} = \frac{\sqrt{1 - p^{t^2}}}{\sqrt{2}}.$$
The density matrix $\rho_{AB}$ (\ref{rhoAB}) can be cast in
 the following matrix form
\begin{eqnarray}
\rho_{AB} =  \frac{2}{{\cal N}_m}\left(
\begin{array}{cccc}
(1+q_r)a^2_{\alpha}a^2_{\alpha t} & 0 & 0 &
(1+q_r)a_{\alpha}a_{\alpha t}b_{\alpha
}b_{\alpha t}\\
0 & (1-q_r)a^2_{\alpha }b^2_{\alpha t} & (1-q_r)a_{\alpha}a_{\alpha
t}b_{\alpha
}b_{\alpha t} & 0 \\
0 & (1-q_r)a_{\alpha }a_{\alpha t}b_{\alpha }b_{\alpha t} &
(1-q_r)b^2_{\alpha }a^2_{\alpha t}
& 0 \\
(1+q_r)a_{\alpha} a_{\alpha t}b_{\alpha }b_{\alpha t} & 0 & 0 &
(1+q_r)b^2_{\alpha }b^2_{\alpha t}
\end{array}
\right)  \label{Xform1}
\end{eqnarray}
in the representation spanned by two-qubit product states
$$|1\rangle = |u_{\alpha } \rangle_A
\otimes |u_{\alpha t} \rangle_B  \quad |2\rangle = | u_{\alpha}
\rangle_A \otimes | v_{\alpha t}\rangle_B \quad |3\rangle = |
v_{\alpha }\rangle_A \otimes |u_{\alpha t}\rangle_B \quad |4\rangle
= |v_{\alpha }\rangle_A \otimes |v_{\alpha t}\rangle_B.$$ In
(\ref{Xform1}), the quantity $q_r$ is defined by
$$ q_r = c_r\cos(m\pi).$$
It is simple to check that the Wootters concurrence \cite{conc}
writes
\begin{equation}\label{CAB}
C(\rho_{AB}) = p^{r^2}\frac{\sqrt{1-p^2}\sqrt{1-p^{2t^2}}}{1 +
p^2\cos m\pi}
\end{equation}
which coincides with the concurrence of the quasi-Bell cat states
$\vert \alpha, \pm \alpha ; m \rangle$ when $t = 1$. It follows that
the bipartite quantum entanglement of formation in the state
$\rho_{AB}$ is
\begin{equation}\label{EAB}
 E(\rho_{AB})= H\bigg(\frac{1}{2} + \frac{1}{2} \frac{\sqrt{1 + 2p^2\cos m\pi+ p^{2r^2}(p^2 - 1)}}{1 +
p^2\cos m\pi}\bigg)
\end{equation}
where $H(x) = - x \log_2 x - (1-x) \log_2 (1 -x)$. Remark that the
reduced density  $\rho_{AE}$ (\ref{rhoAE}) can be obtained from the
density $\rho_{AB}$ (\ref{rhoAB}) by interchanging the roles of the
transmission and reflection parameters $r$ and $t$. Accordingly, the
state $\rho_{AE}$  can be converted in a two qubits system
analogously to qubit mapping realized for the system $AB$. Then, it
is easy to see that the concurrence is
\begin{equation}\label{CAE}
C(\rho_{AE}) = p^{t^2}\frac{\sqrt{1-p^2}\sqrt{1-p^{2r^2}}}{1 +
p^2\cos m\pi},
\end{equation}
and the entanglement of formation writes as
\begin{equation}\label{EAE}
E(\rho_{AE})= H\bigg(\frac{1}{2} + \frac{1}{2}\frac{\sqrt{1 +
2p^2\cos m\pi+ p^{2t^2}(p^2 - 1)}}{1 + p^2\cos m\pi} \bigg).
\end{equation}
Finally, the pure system $ABE$ can be partitioned into two qubits
subsystems $A$ and $BE$. For the first mode $A$, we consider the two
dimensional basis spanned by the vectors $\vert u_{\alpha} \rangle$
and $\vert v_{\alpha} \rangle$ defined by (\ref{qubitA}). For the
subsystem $BE$, we introduce two logical qubits $\vert 0 \rangle$
and $\vert 1 \rangle$ as follows
\begin{equation}
\vert \pm \alpha t , \pm \alpha r \rangle = a_{\alpha} \vert 0
\rangle + b_{\alpha}\vert 1 \rangle \qquad \vert \mp \alpha t , \mp
\alpha r \rangle = a_{\alpha} \vert 0 \rangle - b_{\alpha}\vert 1
\rangle
\end{equation}
where $a_{\alpha}$ and $b_{\alpha}$ are given by
$$a_{\alpha} = \frac{\sqrt{1 + p}}{\sqrt{2}} \quad b_{\alpha} = \frac{\sqrt{1 - p}}{\sqrt{2}}.$$
It follows that, for $ m = 0 ~ ({\rm mod}~ 2)$, we have
\begin{eqnarray}
\rho_{A\vert BE} =  \frac{4}{{\cal N}_0}\left(
\begin{array}{cccc}
a^4_{\alpha} & 0 & 0 & a^2_{\alpha}b^2_{\alpha}\\
0 & 0 & 0 & 0 \\
0 & 0 & 0 & 0  \\
a^2_{\alpha}b^2_{\alpha} & 0 & 0 & b^4_{\alpha }
\end{array}
\right), \label{Xform3}
\end{eqnarray}
and for $m = 1 ~ ({\rm mod} ~2)$, we have
\begin{eqnarray}
\rho_{A\vert BE} =  \frac{4a^2_{\alpha}b^2_{\alpha}}{{\cal
N}_1}\left(
\begin{array}{cccc}
0 & 0 & 0 & 0\\
0 & 1 & 1 & 0 \\
0 & 1 & 1 & 0  \\
0 & 0 & 0 & 0
\end{array}
\right), \label{Xform4}
\end{eqnarray}
in the basis $\{ \vert u_{\alpha}, 0 \rangle, \vert u_{\alpha} , 1
\rangle, \vert v_{\alpha} , 0 \rangle,\vert v_{\alpha} , 1 \rangle
\}$.  In this representation, the concurrence writes
\begin{equation}\label{CABE}
C(\rho_{A\vert BE}) = \frac{1-p^2}{1 + p^2\cos m\pi},
\end{equation}
from which one derives the entanglement of formation
\begin{equation}\label{E-ABE}
E(\rho_{A\vert BE})= H\bigg(\frac{1}{2} + \frac{1}{2}\frac{p\cos
\frac{m\pi}{2}}{1 + p^2\cos m\pi}\bigg).
\end{equation}
As expected, it is completely independent of the reflection and
transmission parameters $r$ and $t$.
\subsection{Monogamy of concurrence and entanglement of formation}
To examine the monogamy relation of entanglement measured by the
concurrence in quantum systems involving three qubits, Coffman et al
\cite{Coffman} introduced the so called three tangle. It is defined
from the bipartite concurrences as
\begin{equation}
\tau_{A,B,E} = C^2(\rho_{A\vert BE}) - C^2(\rho_{A
B})-C^2(\rho_{AE}).
\end{equation}
From equations (\ref{CAB}), (\ref{CAE}) and (\ref{CABE}), we obtain
\begin{equation}
\tau_{A,B,E} = (1-p^2)\frac{(1+p^2)- (p^{2r^2} + p^{2t^2})}{(1 +
p^2\cos m\pi)^2}.
\end{equation}
In the figures 1 and 2, corresponding respectively to symmetric and
antisymmetric quasi-Bell cat states, we plot the three tangle
$\tau_{A,B,E}$ as a function of $p$ and $t^2$ . As it can be easily
seen, $\tau_{A,B,E}$ is always positive. The inequality given by
(\ref{mono-relation}) is then satisfied. This indicates that the
 squared concurrence is a monogamous measure.
\begin{center}
  \includegraphics[width=3in]{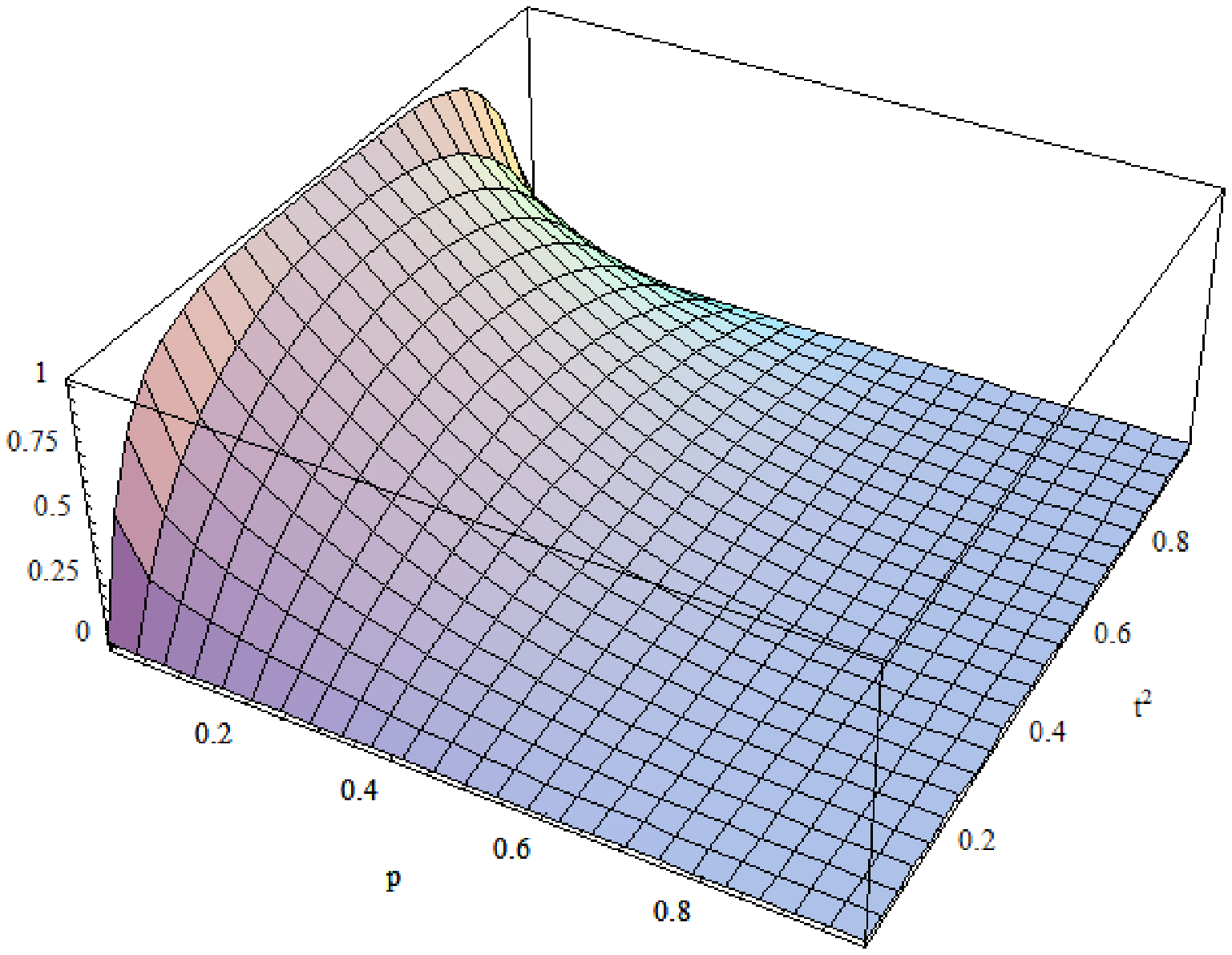}\\
{\bf Figure 1.}  {\sf The three tangle  $\tau_{A,B,E}$ versus the
overlapping $p$ and the transmission $t^2$ for  $m=0$.}
\end{center}
\begin{center}
  \includegraphics[width=3in]{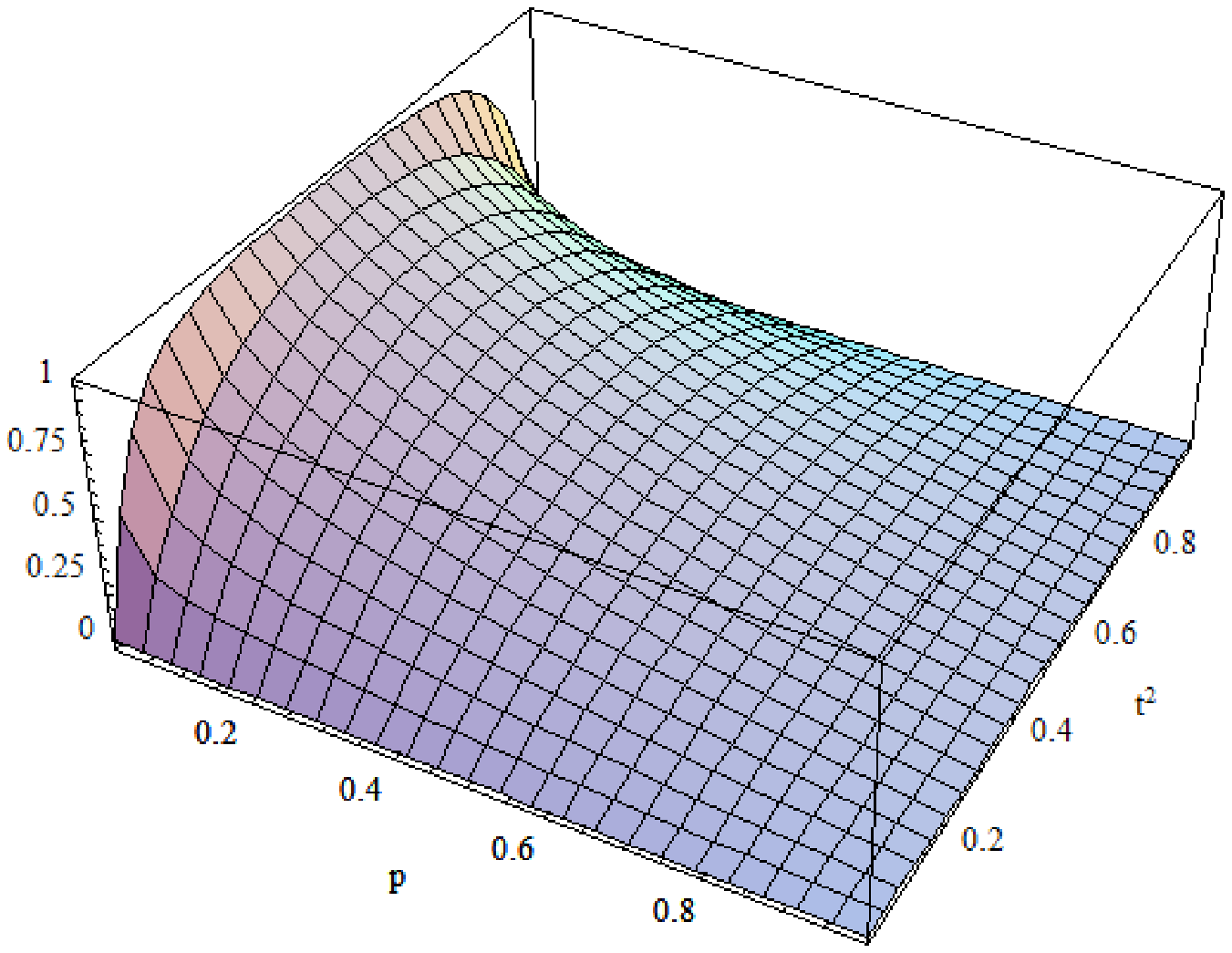}\\
{\bf Figure 2.}  {\sf The three tangle  $\tau_{A,B,E}$ versus the
overlapping $p$ and the transmission $t^2$ for  $m=1$.}
\end{center}
In particular when the decoherence effect is modeled by a 50:50 beam
splitter (i.e. $r^2 = t^2 = {1\over 2}$), one obtains
$$\tau_{A,B,E}(m=0) = \frac{(1-p^2)(1- p)^2}{(1 +
p^2)^2}$$ for symmetric quasi-Bell cat states and for the
antisymmetric ones one has
$$\tau_{A,B,E}(m=1) = \frac{ 1-p}{1 +
p}$$
Clearly, for both cases, $\tau_{A,B,E}$ is positive reflecting the
monogamous property of the squared concurrence in agrement with the
results reported in the figures 1 and 2. In the limiting case, $p
\longrightarrow 0$ (resp. $p \longrightarrow 1$), using the
equations (\ref{CAB}), (\ref{CAE}) and (\ref{CABE}), one can check
that $\tau_{A,B,E} = 1$ (resp. $\tau_{A,B,E} = 0$). Similarly, to
decide about the monogamy of entanglement of formation,  we examine
the positivity of the following quantity
\begin{equation}\label{deltaABE}
{\it E}_{A,B,E}\equiv  {\it E}_{A,B,E} (t^2,p)= E(\rho_{A\vert BE})
- E(\rho_{A B})-E(\rho_{AE})
\end{equation}
defined in terms of the bipartite entanglement of formation given by
the equations (\ref{EAB}), (\ref{EAE}) and (\ref{E-ABE}). Noticing
that
$${\it E}_{A,B,E} (t^2,p)= {\it E}_{A,B,E} (r^2=1-t^2,p),$$
we shall restrict our discussion in what follows to the interval $0
\leq t^2 \leq 0.5$.  The behavior of the function $E = E_{A,B,E}$
defined by (\ref{deltaABE}) versus the overlapping $p$ and the
transmission coefficient $t$ is plotted in the figures 3 for even
quasi-Bell cat states $(m=0)$. It is symmetric with respect to the
$t^2 = \frac{1}{2}$-axis as expected. The figure 3.a shows that the
function $E_{A,B,E}$ is not always positive for symmetric quasi-Bell
cat states and the entanglement of formation does not satisfy the
monogamy relation (\ref{mono-relation}) for small values of $t^2$.
To see clearly this feature, we plot in the figure 3.b, the quantity
$E_{A,B,E}$ for transmission $t^2$ ranging from 0.0125 to 0.2. The
figure 3.b reveals that for $t^2 \leq 0.025$, the monogamy relation
is violated for quasi-Bell cat states involving Glauber coherent
states with overlap such that  $0 \leq p \leq 0.4$.
\begin{center}
  \includegraphics[width=3in]{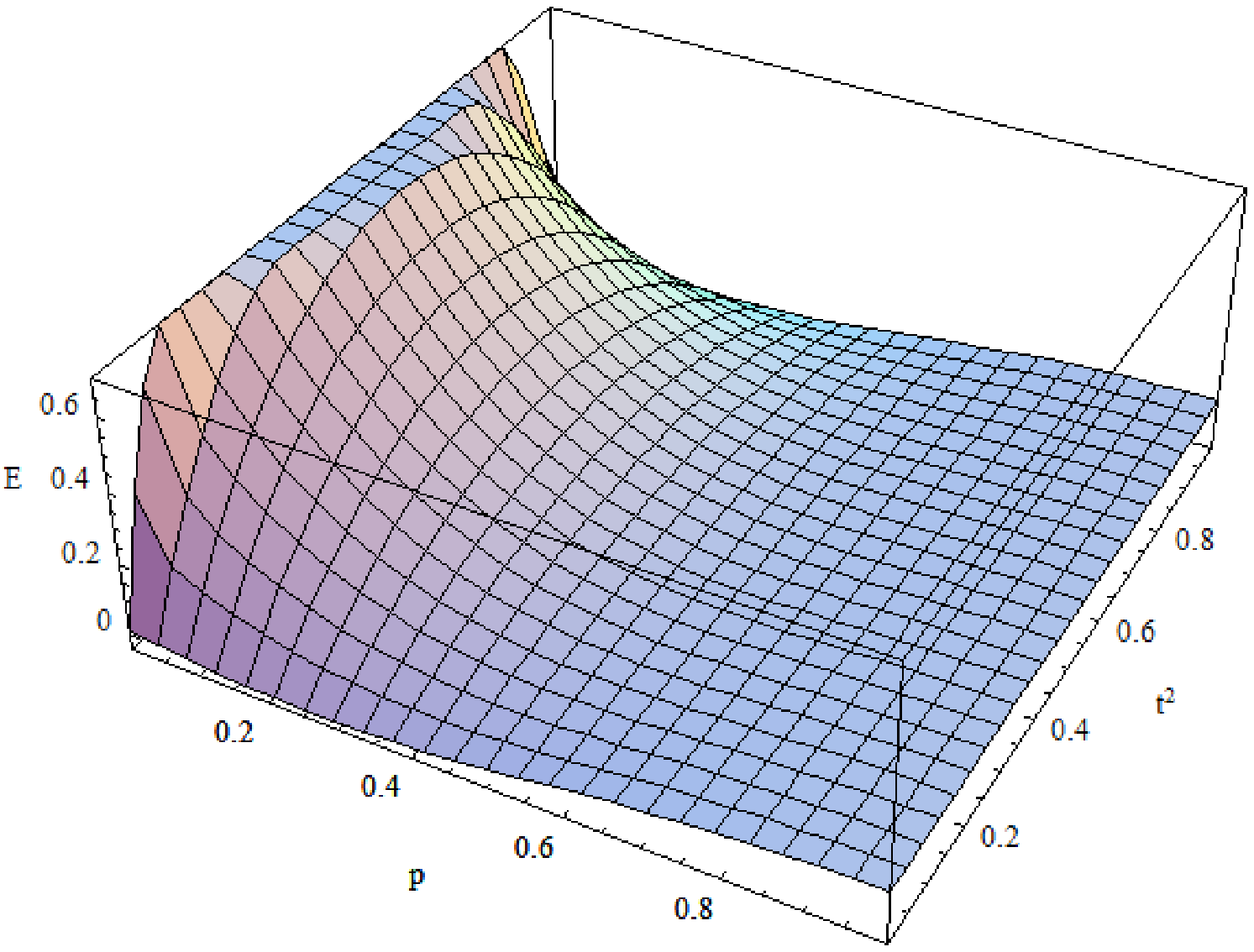}\\
{\bf Figure 3.a.}  {\sf $E = E_{A,B,E}$ versus the overlapping $p$
and the transmission $t^2$ for  $m=0$.}
\end{center}


\begin{center}
  \includegraphics[width=3in]{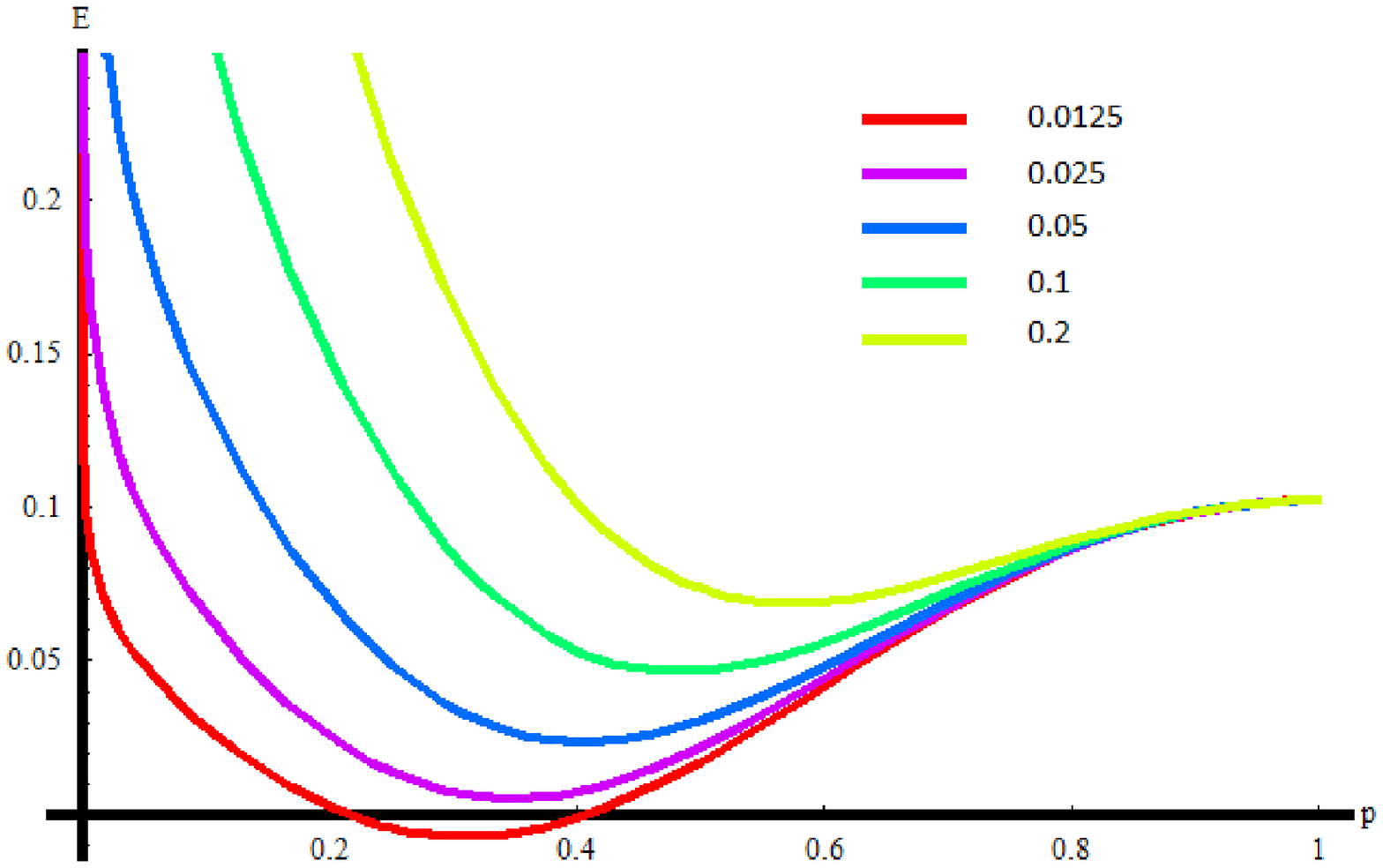}\\
{\bf Figure 3.b}  {\sf  $E = E_{A,B,E}$ versus the overlapping $p$
for small values of $t^2$ when $m=0$.}
\end{center}
For odd quasi-Bell cat states $(m=1)$,  the function $E = E_{A,B,E}$
vs the transmission coefficient $t^2$ and the overlap $p$ is
reported in the figures 4. The plot shows that the function $E =
E_{A,B,E}$ decreases quickly from the unity to vanishes when $p
\simeq 0.33$ for any value of the transmission parameter $t^2$. It
follows that for $ 0 \leq p \leq 0.33 $ the entanglement of
formation is monogamous. The function $E = E_{A,B,E}$ becomes
negative and the monogamy inequality cease to be satisfied for $
0.33 \leq p \leq 1$.


\begin{center}
\includegraphics[width=6in]{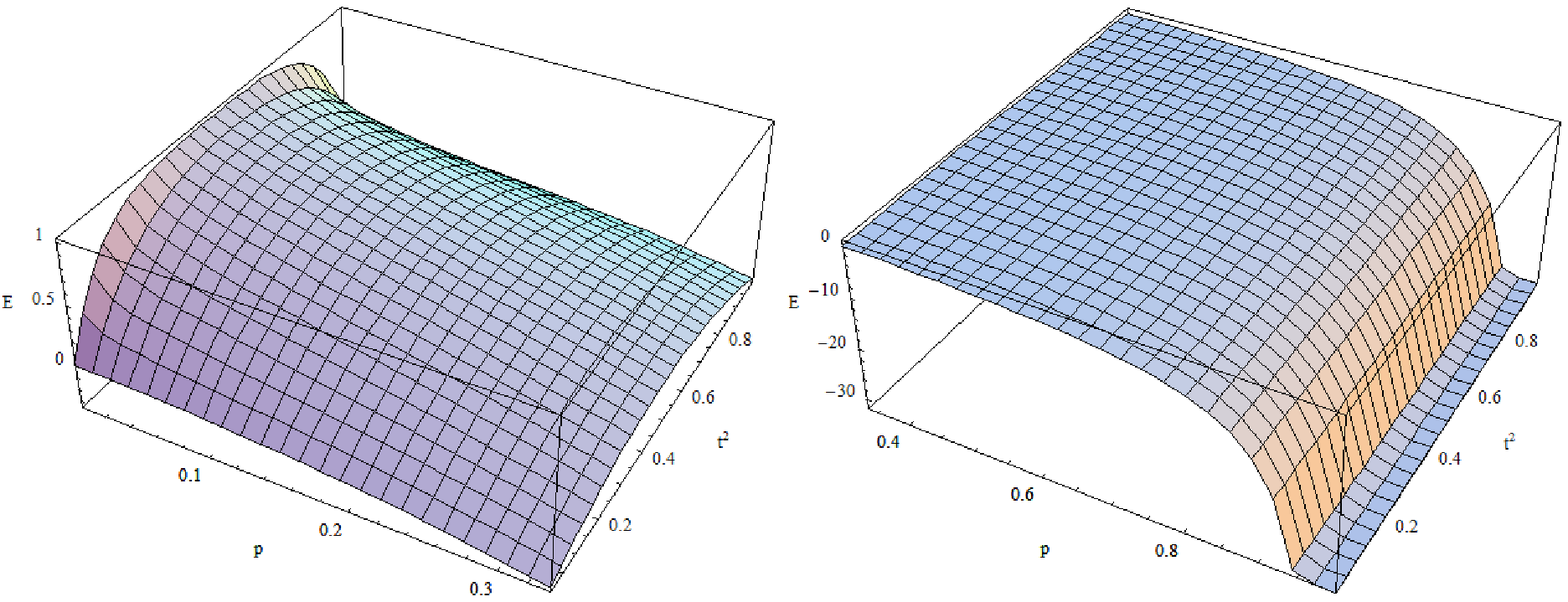}\\
{\bf Figure 4.}  {\sf   $E = E_{A,B,E}$ versus the overlapping $p$
and the transmission $t^2$ for  $m=1$..}
\end{center}



\section{Quantum discord}

\subsection{Definition and Koashi-Winter relation }
For a state $\rho_{AB}$  of a bipartite quantum system composed of
two  particles or modes  $A$ and  $B$, the quantum discord is
defined as the difference between total correlation $I(\rho_{AB})$
and classical correlation $J(\rho_{AB})$. The total correlation is
usually quantified by the mutual information:
\begin{equation}
    I(\rho_{AB})=S(\rho_A)+S(\rho_B)-S(\rho_{AB}),
\end{equation}
where $\rho_{A(B)}={\rm Tr}_{B(A)}(\rho_{AB})$ is the reduced state
of $A$($B$), and $S(\rho)$ is the von Neumann entropy of a quantum
state $\rho$. Suppose that a positive operator valued measure (POVM)
measurement is performed on particle $A$. The set of POVM elements
is denoted by $\mathcal{M}=\{M_k\}$ with $M_k\geqslant 0$ and
$\sum_k M_k= \mathbb{I} $. We remind that the generalized positive
operator valued measurement is not required. Indeed, it has be shown
in \cite{Hamieh} that  the optimal measurement of the conditional
entropy is ensured by projective one. Thus, a projective measurement
on the subsystem $A$ project the system into a statistical ensemble
$\{ p^B_k , \rho_{B_{k}}\}$, such that
\begin{eqnarray}
\rho_{AB} \longrightarrow \rho_{B_{k}} = \frac{(M_k \otimes
\mathbb{I})\rho_{AB}(M_k \otimes \mathbb{I})}{p^B_k}\label{rhoBk}
\end{eqnarray}
where the von Neumann measurement for subsystem $A$ writes as
\begin{eqnarray}
M_k = U \, \Pi_k \, U^\dagger : \quad k = 0,1 \, , \label{Eq:VNmsur}
\end{eqnarray}
with $\Pi_k = |k\rangle\langle k|$ is the projector for subsystem
$A$  along the computational base $|k\rangle$,  $U \in SU(2)$ is a
unitary operator with unit determinant, and
$$ p^B_k = {\rm tr}  \bigg[ (M_k \otimes \mathbb{I})\rho_{AB}(M_k \otimes \mathbb{I}) \bigg].$$
The classical correlation is then obtained by performing the
maximization over all the measurements. This gives
\begin{eqnarray}
J(\rho_{AB})& =\max_{\mathcal{M}}
\Big[S(\rho_B)-\sum_k ~p^B_k ~S(\rho_{B_{k}})\Big] \nonumber \\
& =S(\rho_B) - \widetilde{S}_{\rm min} \label{def: classical
correlation}
\end{eqnarray}
where $\widetilde{S}_{\rm min}$  denotes the minimal value of the
conditional  entropy
\begin{equation}
\widetilde{S} =  \sum_k ~p^B_k
~S(\rho_{B_{k}}).\label{condit-entropy}
\end{equation}
Then, the difference between $I(\rho_{AB})$ and $J(\rho_{AB})$ gives
the amount of quantum discord present in the bipartite system $AB$
\begin{equation} \label{def: discord}
    D(\rho_{AB})= I(\rho_{AB}) - J(\rho_{AB})
    =S(\rho_A)+\widetilde{S}_{\rm min}-S(\rho_{AB}).
\end{equation}
The minimal value of the conditional entropy is related to the
entanglement of formation of $E(\rho_{BC})$ of the state $\rho_{BC}$
which the complement of the density $\rho_{AB}$ . This relation is
the so-called Koashi-Winter relation \cite{Koachi-Winter}. It is
given by
\begin{equation}\label{Smin}
\widetilde{S}_{\rm min} = E(\rho_{BC}) = H\bigg(\frac{1}{2} +
\frac{1}{2} \sqrt{1 - \vert C(\rho_{BC})\vert^2}\bigg)
\end{equation}
and establishes the connection between the classical correlation of
a bipartite state $\rho_{AB}$ and the entanglement of formation of
its complement $\rho_{BC}$. This connection requires a purification
of the state $\rho_{AB}$ by an ancilla qubit C and provides an
explicit algorithm to determine  the quantum discord especially for
rank-two states.
\subsection{Analytical computation of quantum discord}
To evaluate the quantum discord present in the state (\ref{rhoAB}),
we first calculate the mutual information $I (\rho_{AB})$. The
density $\rho_{AB}$ is a two qubit state of rank two. The
corresponding non vanishing  eigenvalues  are given by
\begin{eqnarray}
\lambda^{AB}_{\pm} =  \frac{(1\pm p^{r^2}\cos m\pi)(1 \pm
p^{t^2+1})}{2 + 2p^2\cos m\pi}. \label{lambda12}
\end{eqnarray}
It follows that the joint entropy is
\begin{eqnarray}
S(\rho_{AB}) =  - \lambda^{AB}_{+} \log_2\lambda^{AB}_{+} -
\lambda^{AB}_{-} \log_2\lambda^{AB}_{-}. \label{SAB}
\end{eqnarray}
The quantum mutual information is then given by
\begin{eqnarray}
I (\rho_{AB}) = S(\rho_A) + S(\rho_B) + \sum_{j=+,-}
\lambda^{AB}_{j} \log_2 \lambda^{AB}_{j} \label{IAB}
\end{eqnarray}
where $\rho_A$ and $\rho_B$ are the marginal states of $\rho_{AB}$,
and
\begin{eqnarray}
S(\rho_A) =  - \lambda^A_+ \log_2\lambda^A_+ - \lambda^A_- \log_2
\lambda^A_- \qquad S(\rho_B) =  - \lambda^B_+ \log_2\lambda^B_+ -
\lambda^B_- \log_2 \lambda^B_- \label{SA+SB}
\end{eqnarray}
with
$$ \lambda^A_{\pm} = \frac{1}{2} (1 \pm p) \frac{1\pm p\cos m\pi}{1 +  p^2\cos m\pi}
\qquad \lambda^B_{\pm} = \frac{1}{2} (1 \pm p^{t^2})\frac{1\pm
p^{r^2+1}\cos m\pi}{2 + 2 p^2\cos m\pi}.$$ Reporting (\ref{SA+SB})
into (\ref{IAB}), the quantum mutual information reads
\begin{eqnarray}
I(\rho_{AB})=  H(\lambda^A_{+}) + H(\lambda^B_{+})-
H(\lambda^{AB}_{+})
\end{eqnarray}
To derive the explicit form of the classical correlation
$J(\rho_{AB})$, we decompose the state $\rho_{AB}$ as
\begin{eqnarray}
\rho_{AB} = \lambda^{AB}_{+} \vert \psi_+ \rangle \langle \psi_+
\vert + \lambda^{AB}_{-} \vert \psi_- \rangle \langle \psi_- \vert
\end{eqnarray}
where the eigenvalues $\lambda^{AB}_{\pm}$ are given by
(\ref{lambda12}) and the eigenstates $\vert \psi_{\pm} \rangle$  are
\begin{eqnarray}
\vert \psi_+ \rangle = \frac{1} {\sqrt{a^2_{\alpha}a^2_{\alpha t} +
b^2_{\alpha}b^2_{\alpha t}}} ( a_{\alpha}a_{\alpha t} \vert
u_{\alpha},u_{\alpha t} \rangle + b_{\alpha}b_{\alpha t} \vert
v_{\alpha},v_{\alpha t} \rangle)
 \nonumber \\
\vert \psi_- \rangle = \frac{1} {\sqrt{a^2_{\alpha}b^2_{\alpha t} +
b^2_{\alpha}a^2_{\alpha t}}} ( a_{\alpha}b_{\alpha t} \vert
u_{\alpha}, v_{\alpha t} \rangle + b_{\alpha}a_{\alpha t} \vert
v_{\alpha}, v_{\alpha t} \rangle).
\end{eqnarray}
Attaching a qubit $C$ to the bipartite system $AB$, we write the
purification of $\rho_{AB}$ as
\begin{eqnarray}
\vert \psi \rangle = \sqrt{\lambda^{AB}_+} \vert \psi_+ \rangle
\otimes \vert u_{\alpha} \rangle +  \sqrt{\lambda^{AB}_-} \vert
\psi_- \rangle \otimes \vert v_{\alpha} \rangle
\end{eqnarray}
such that the whole system $ABC$ is described by the pure density
state $\rho_{ABC} = \vert \psi \rangle \langle \psi \vert $ so that
$\rho_{AB} = {\rm Tr}_C \rho_{ABC}$ and $\rho_{BC} = {\rm Tr}_A
\rho_{ABC}$. As mentioned above, The Koachi and Winter relation
\cite{Koachi-Winter} simplifies drastically the minimization process
of the conditional entropy and  the minimal amount of conditional
entropy coincides with the entanglement of formation of $\rho_{BC}$.
Therefore, employing the prescription presented in \cite{conc}, the
entanglement of formation in the state  $\rho_{BC}$ is
\begin{equation}
\widetilde{S}_{\rm min} = E(\rho_{BC}) = H\bigg(\frac{1}{2} +
\frac{1}{2} \sqrt{1 - \vert C(\rho_{BC})\vert^2}\bigg)\label{Smin}
\end{equation}
with
$$ C(\rho_{BC}) = \frac{\sqrt{p^{2}(1 - p^{2r^2})(1 - p^{2t^2})}}{(1+p^{2}\cos m\pi)}.$$
It must be noticed that this result can be alternatively obtained
using the minimization procedure presented in \cite{Ali} (see also
\cite{X-M Lu}). According to the equation (\ref{def: classical
correlation}), the classical correlation is
\begin{equation}\label{cc-formulaAB}
J(\rho_{AB}) = H \bigg(\frac{1}{4} (1 + p^{t^2})\frac{1 +
p^{r^2+1}\cos m\pi}{1 +  p^2\cos m\pi}\bigg) - H\bigg(\frac{1}{2} +
\frac{1}{2} \sqrt{1 - p^{2}\frac{(1 - p^{2r^2})(1 -
p^{2t^2})}{(1+p^{2}\cos m\pi)^2}}\bigg)
\end{equation}
and using the definition (\ref{def: discord}), the explicit
expression of quantum discord reads
\begin{equation}\label{qd-formulaAB}
D(\rho_{AB}) = H\bigg( \frac{1 + p}{2}  \frac{1 + p\cos m\pi}{1 +
p^2\cos m\pi} \bigg) - H\bigg( \frac{(1 + p^{r^2}\cos m\pi)(1 +
p^{t^2+1})}{2 + 2p^2\cos m\pi} \bigg) + H\bigg(\frac{1}{2} +
\frac{1}{2} \sqrt{1 - p^{2}\frac{(1 - p^{2r^2})(1 -
p^{2t^2})}{(1+p^{2}\cos m\pi)^2}}\bigg)
\end{equation}
Note that for $r = 0$, the density state $\rho_{AB}$ (\ref{rhoAB})
reduces to the pure density of quasi-Bell cat states (\ref{bell-cs})
and the the quantum discord (\ref{qd-formulaAB}) gives
\begin{equation}
D(\vert \alpha,\pm \alpha; m \rangle) = H\bigg( \frac{1 + p}{2}
\frac{1 + p\cos m\pi}{1 + p^2\cos m\pi} \bigg)
\end{equation}
which coincides with the entanglement of formation of quasi-Bell cat
states given by
$$ E(\vert \alpha,\pm \alpha; m \rangle ) = H\bigg(\frac{1}{2} + \frac{1}{2}
\sqrt{1 - \vert C(\vert \alpha,\pm \alpha; m
\rangle)\vert^2}\bigg)$$ where the concurrence is given by
$$C(\vert \alpha,\pm \alpha; m \rangle ) = \frac{1 - p^2}{1 + p^2 \cos m\pi}.$$
The quantum discord present in the bipartite state $\rho_{AE}$ can
be simply  obtained  from the equation (\ref{qd-formulaAB}) by
interchanging $r$ and $t$. So, we have
\begin{equation}\label{qd-formulaAE}
D(\rho_{AE}) = H\bigg( \frac{1 + p}{2}  \frac{1 + p\cos m\pi}{1 +
p^2\cos m\pi} \bigg) - H\bigg( \frac{(1 + p^{t^2}\cos m\pi)(1 +
p^{r^2+1})}{2 + 2p^2\cos m\pi} \bigg) + H\bigg(\frac{1}{2} +
\frac{1}{2} \sqrt{1 - p^{2}\frac{(1 - p^{2r^2})(1 -
p^{2t^2})}{(1+p^{2}\cos m\pi)^2}}\bigg)
\end{equation}
The state $\rho_{A\vert BE}$ is pure and the quantum discord
coincides with the entanglement of formation
$$D(\rho_{A\vert BE}) = E(\rho_{A\vert BE})$$
given by the expression (\ref{E-ABE}).
\subsection{Monogamy of quantum discord}
Analogously to the treatment of squared concurrence and entanglement
of formation presented in the previous section, we define the
quantity
$$ D_{A,B,E} = D(\rho_{A\vert BE}) - D(\rho_{AB}) - D(\rho_{AE})$$
as the difference between the quantum discord $D(\rho_{A\vert BE})$
and the sum $D(\rho_{AB}) + D(\rho_{AE})$.\\
For symmetric quasi-Bell cat states $(m=0)$,  the numerical results
reported in the figures 5 and 6 show that the quantum discord is
monogamous for any value of the reflection parameter $r$ and the
overlap $p$. For a 50:50 beam splitter, the behavior of the quantity
$D_{A,B,E}$ is given in the figure 6. it reveals that $D_{A,B,E}$,
which is maximal  for $p = 0$, decreases to reach a minimal value
for $p \simeq 0.5$ and increases after slowly.

\begin{center}
  \includegraphics[width=3in]{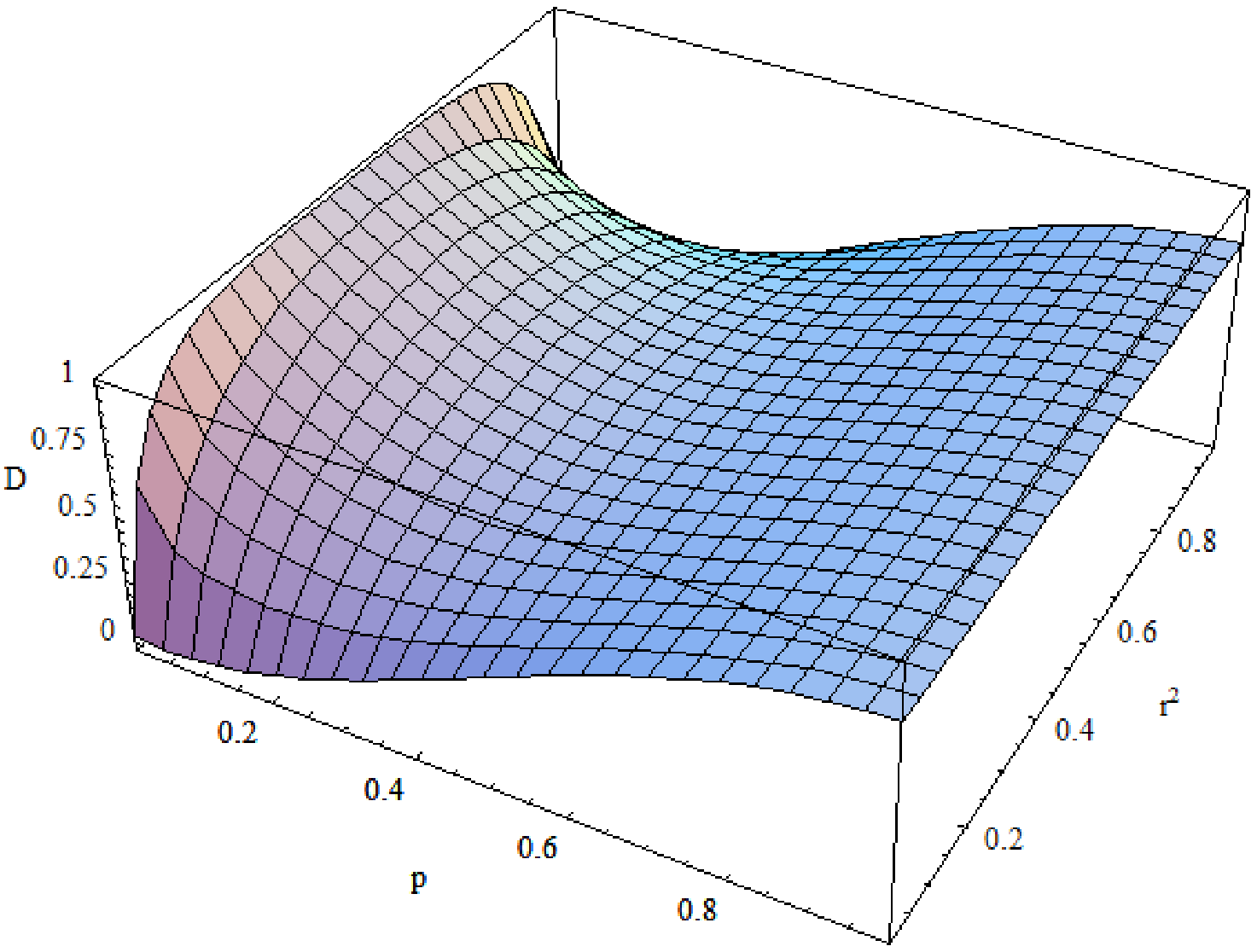}\\
{\bf Figure 5.}  {\sf $D = D_{A,B,E}$ versus the overlapping $p$ and
the transmission $r^2$ for  $m=0$.}
\end{center}
\begin{center}
  \includegraphics[width=3in]{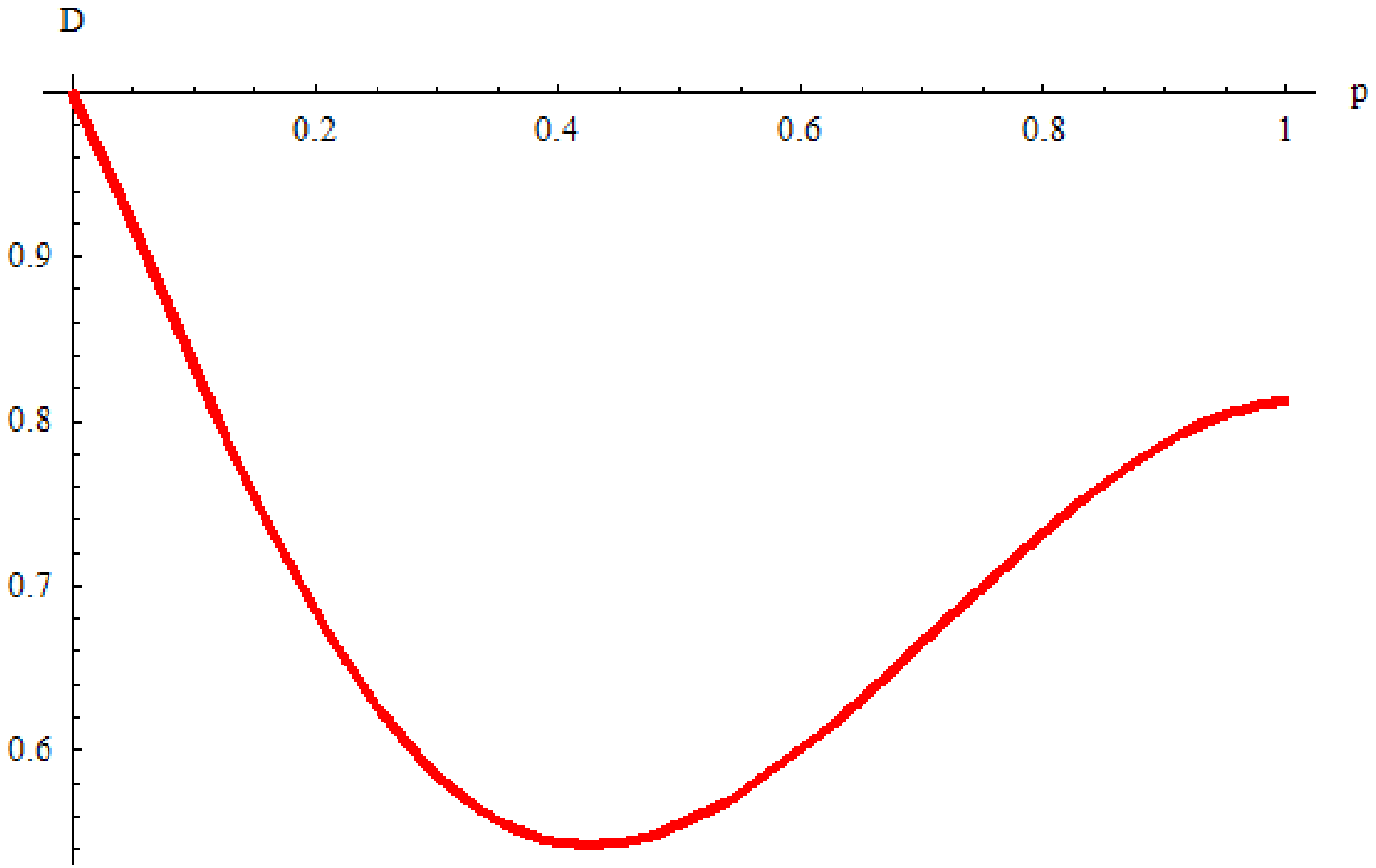}\\
{\bf Figure 6.}  {\sf  $D = D_{A,B,E}$ versus the overlapping $p$
for $t^2= \frac{1}{2}$ and   $m=0$.}
\end{center}
For antisymmetric quasi-Bell cat states $(m = 1)$, the monogamy
becomes violated when the overlap $p$ approaches the unity (see the
figures 7 and 8). This feature is clearly illustrated in the figure
8 corresponding to the situation where $t^2 = 1/2$. The function $D
= D_{A,B,E}$ becomes negative for $0.85 \leq p \leq 1$.

\begin{center}
  \includegraphics[width=3in]{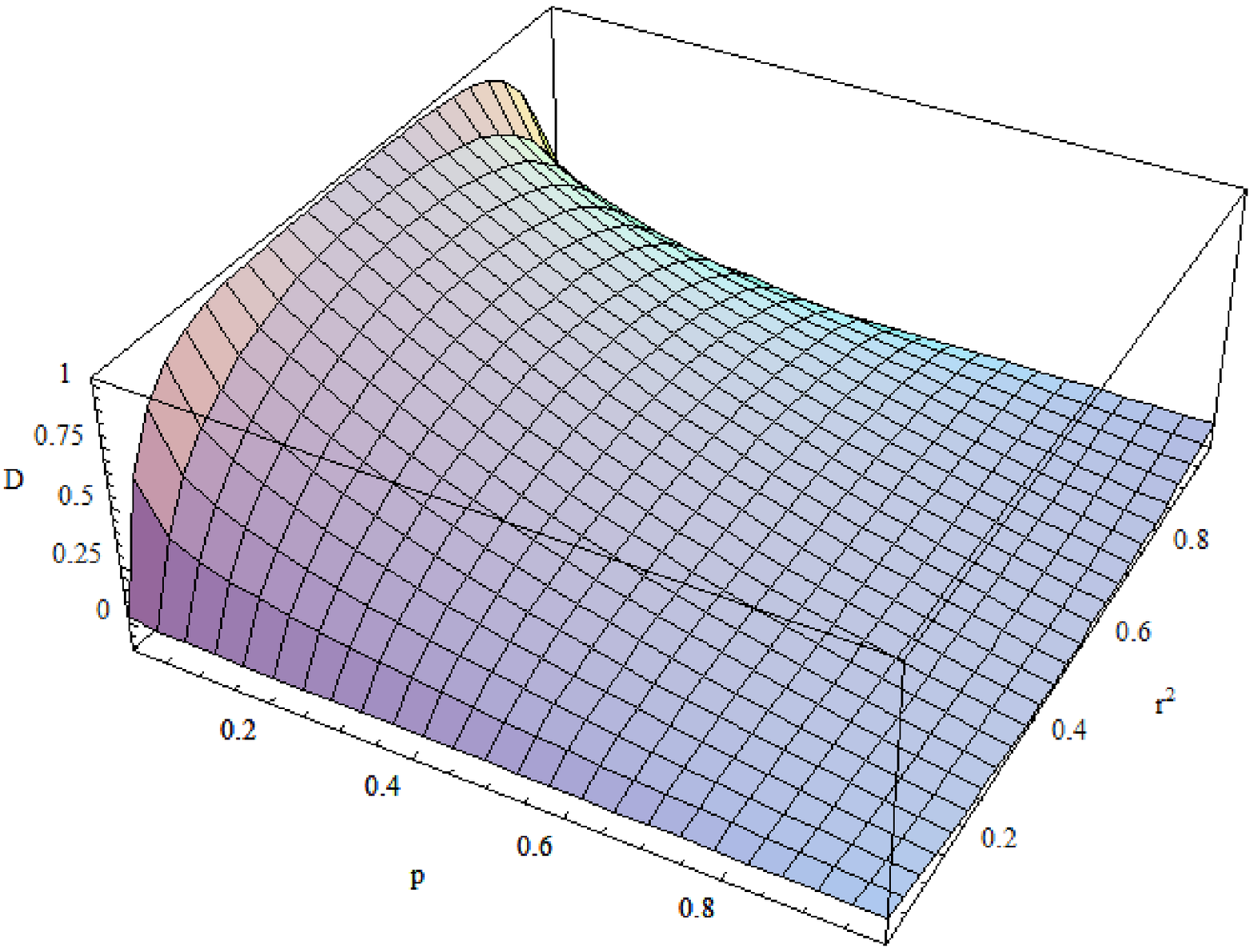}\\
{\bf Figure 7.}  {\sf  $D = D_{A,B,E}$ versus the overlapping $p$
and the transmission $r^2$ for  $m=1$.}
\end{center}

\begin{center}
  \includegraphics[width=3in]{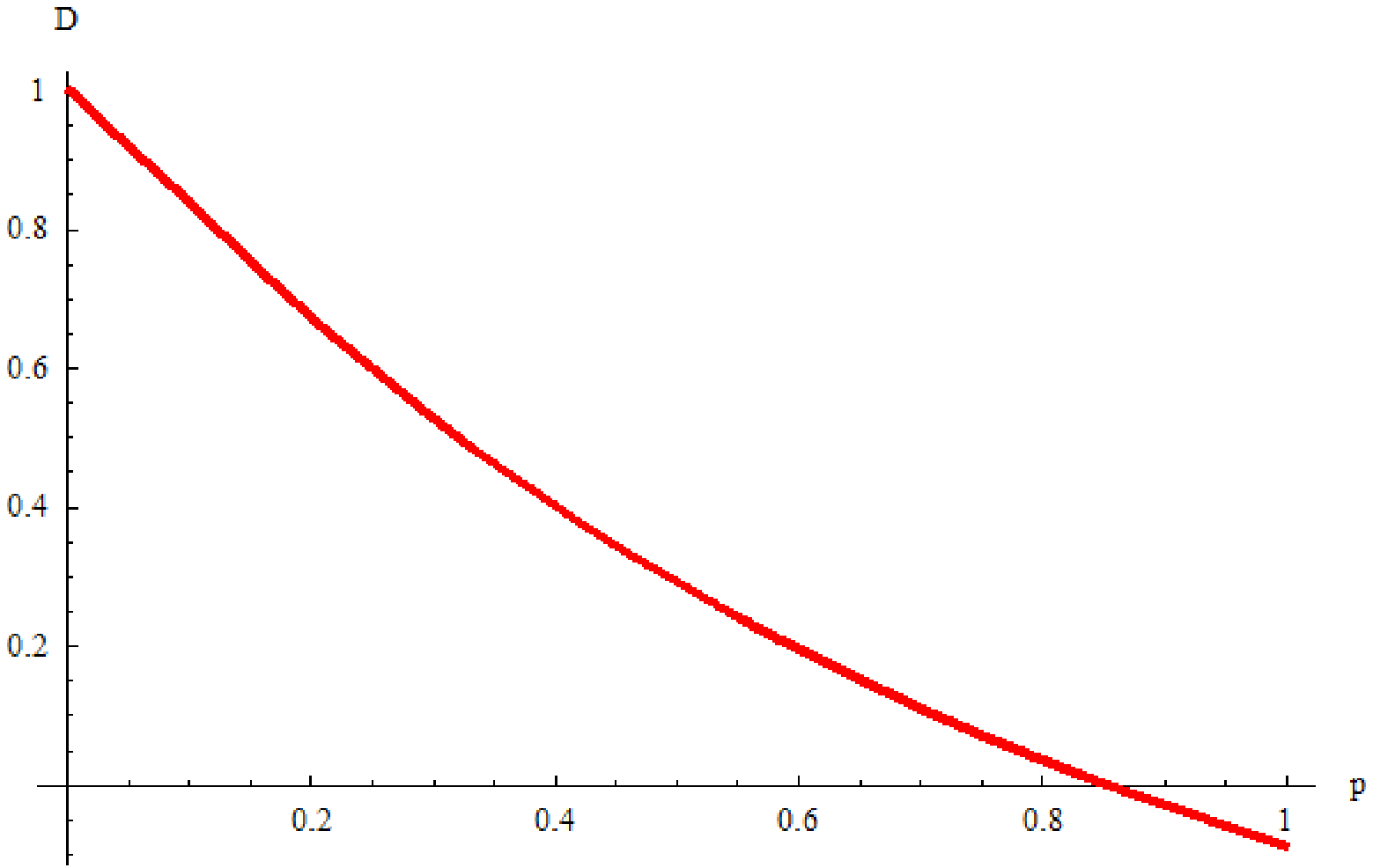}\\
{\bf Figure 8.}  {\sf  $D = D_{A,B,E}$ versus the overlapping $p$
for $t^2= \frac{1}{2}$ and   $m=1$.}
\end{center}

\section{Geometric measure of quantum discord}

\subsection{Geometric quantum discord: Generalities}
The geometrized version of quantum discord, introduced by  Dakic et
al  \cite{Dakic2010}, measures the distance between a state $\rho$
of a bipartite system $AB$ and the closest classical-quantum state
presenting zero discord. It is defined by
\begin{equation}
  D_{g}(\rho):=\min_{\chi}||\rho-\chi||^{2}
\end{equation}
where the minimum is over the set of zero-discord states $\chi$ and
the distance is the square norm in the Hilbert-Schmidt space:
$$||\rho-\chi||^{2}:= {\rm Tr}(\rho-\chi)^2. $$
When the measurement is taken on the subsystem $A$, the zero-discord
state $\chi$ can be represented as \cite{Ollivier-PRL88-2001}
$$\chi= \sum_{i = 1,2}p_{i}|\psi_{i}\rangle\langle\psi_{i}|\otimes\rho_{i}
$$
where $p_i$ is a probability distribution, $\rho_{i}$ is the
marginal density matrix of $B$ and  $\{|\psi_1\rangle
,|\psi_2\rangle \}$ is an arbitrary orthonormal vector set. A
general two qubit state writes in Bloch representation as
\begin{eqnarray}
  \rho & = & \frac{1}{4}\left[ \sigma_{0}\otimes \sigma_{0} +\sum_{i}^{3}(x_{i}\sigma_{i}\otimes \sigma_{0}
   +y_{i} \sigma_{0}\otimes\sigma_{i})+\sum_{i,j=1}^{3}R_{ij}\sigma_{i}\otimes\sigma_{j}\right]
\end{eqnarray}
where $x_{i} = {\rm Tr}\rho(\sigma_{i}\otimes \sigma_{0}),~  y_{i} =
{\rm Tr}\rho(\sigma_{0}\otimes\sigma_{i})$ are components of local
Bloch vectors and $R_{ij} = {\rm
Tr}\rho(\sigma_{i}\otimes\sigma_{j})$ are components of the
correlation tensor. The operators $\sigma_i$ $( i = 1, 2, 3)$ stand
for the three Pauli matrices and $\sigma_0$ is the identity matrix.
The explicit expression of the geometric measure of quantum discord
is given by \cite{Dakic2010}:
\begin{equation}
  D_g(\rho)=\frac{1}{4}\left(||x||^{2}+||R||^{2}-k_{\rm {max}}\right) \label{eq:GMQD_original}
\end{equation}
where $x=(x_{1},x_{2},x_{3})^{T}$, $R$ is the matrix with elements
$R_{ij}$, and $k_{\rm{max}}$ is the largest eigenvalue of matrix
defined by
\begin{equation} \label{matrix K}
K := xx^{T}+RR^{T}.
\end{equation}
Denoting the eigenvalues of the $3\times 3$ matrix $K$ by
$\lambda_1$, $\lambda_2$ and $\lambda_3$ and considering
$||x||^{2}+||R||^{2}={\rm Tr}K$, we get an alternative compact form
of the geometric measure of quantum discord
\begin{equation}
D_g(\rho) = \frac{1}{4}~ {\rm min}\{ \lambda_1 + \lambda_2 ,
\lambda_1 + \lambda_3 , \lambda_2 + \lambda_3\}.\label{eq:GMQD_new}
\end{equation}

\subsection{Explicit expressions}
The density $\rho ^{AB}$  (\ref{rhoAB}) writes, in the Bloch
representation, as
\begin{equation}
\rho_{AB}=\frac{1}{4}\bigg(\sigma _{0}\otimes \sigma _{0}+ R_{30}
~\sigma _{3}\otimes \sigma _{0} + R_{03} ~\sigma _{0}\otimes \sigma
_{3}+ \sum_{i=1}^{3} R_{ii}~ \sigma _{i}\otimes \sigma _{i}\bigg)
\end{equation}
where the correlation matrix elements are given by
\begin{equation}
R_{03}=\frac{p^{t^2} + p^{2-t^2}\cos m\pi}{1+p^2\cos m\pi } {\hskip
1cm} R_{30}=\frac{p(1+\cos m\pi)}{1+p^2\cos m\pi }
\end{equation}
\begin{eqnarray}
R_{11} = \frac{\sqrt{(1-p^{2})(1-p^{2t^{2}})}}{1+p^2\cos m\pi
}{\hskip 0.5cm} R_{22}=-p^{r^{2}}\cos m\pi
\frac{\sqrt{(1-p^{2})(1-p^{2t^{2}})}}{ 1+p^2\cos m\pi } {\hskip
0.5cm} R_{33} = \frac{p^{1-t^{2}}\cos m\pi +p^{1+t^{2}}}{1+p^2\cos
m\pi }.
\end{eqnarray}
The eigenvalues of the matrix $K$ , defined by (\ref{matrix K}), are
thus given
$$
\lambda _{1}=R_{30}^{2}+R_{33}^{2} \qquad\qquad \lambda
_{2}=R_{11}^{2} \qquad\qquad \lambda _{3}=R_{22}^{2}
$$
in terms of the elements of the matrix correlation. They also
rewrite as
\begin{eqnarray}
\lambda _{1} = p^2~\frac{p^{2t^2} + p^{-2t^2} + 4\cos m\pi + 2}{(1 +
p^2 \cos m\pi)^2},\quad \lambda _{2} =
\frac{(1-p^{2})(1-p^{2t^{2}})}{(1 + p^2 \cos m\pi)^2},\quad \lambda
_{3} = p^{2r^2}\frac{(1-p^{2})(1-p^{2t^{2}})}{(1 + p^2 \cos m\pi)^2}
\end{eqnarray}
It is clear that $\lambda_3 \leq \lambda_2$ and we have
\begin{equation}
D_{g}(\rho _{AB})=\frac{1}{4}\min \{\lambda _{1}+\lambda
_{3},\lambda _{2}+\lambda _{3}\}
\end{equation}
For  $\lambda _{1} \geq \lambda _{2}$, the geometric measure of
quantum discord gives
\begin{equation}
D_{g}=\frac{\lambda _{2}+\lambda _{3}}{4}.
\end{equation}
Alternatively, for $\lambda _{1} \leq \lambda _{2}$, one obtains
\begin{equation}
D_{g}=\frac{\lambda _{1}+\lambda _{3}}{4}.
\end{equation}
Explicitly, the condition $\lambda _{1} \geq \lambda _{2}$ writes as
\begin{equation}\label{conditionmaster}
p^{2r^2}+ p^{2t^2} + p^2(4\cos m\pi + 3) - 1 \geq 0.
\end{equation}
A condition that we shall discuss separately for the symmetric  and
anti-symmetric cases. We first consider the situation where $m = 0$.
In this case, the condition (\ref{conditionmaster}) becomes
\begin{equation}\label{condition0}
p^{2r^2}+ p^{2t^2} + 7 p^2 - 1 \geq 0.
\end{equation}
which is satisfied  when $ \frac{2\sqrt{2}-1}{7} \leq p \leq 1$ for
all possible values of $t$ ranging between 0 and 1. It follows that,
for $ \frac{2\sqrt{2}-1}{7} \leq p \leq 1$, the geometric discord is
given by
\begin{equation}
D_{g}(\rho _{AB})=\frac{\lambda _{2}+\lambda _{3}}{4} =
\frac{1+p^{2r^2}}{4}\frac{(1-p^{2})(1-p^{2t^{2}})}{(1 + p^2)^2}.
\end{equation}
For $0 \leq p \leq \frac{2\sqrt{2}-1}{7}$, the condition
(\ref{condition0}) is satisfied for
$$ 0 \leq t^2 \leq t_-^2 \qquad  t_+^2 \leq t^2 \leq 1 $$
where
$$ t_{\mp}^2 = \frac{1}{2} + \frac{1}{2} \frac{\ln \bigg[ \frac{1-7p^2}{2p} \pm \sqrt{\bigg(\frac{1-7p^2}{2p}\bigg)^2 -1}\bigg]}{\ln p}.$$
In this situation, the geometric quantum discord is
\begin{equation}
D_{g}(\rho _{AB})=\frac{\lambda _{2}+\lambda _{3}}{4} =
\frac{1+p^{2r^2}}{4}\frac{(1-p^{2})(1-p^{2t^{2}})}{(1 + p^2)^2}.
\end{equation}
However, for coherent states with overlapping $p$ such that $0 \leq
p \leq \frac{2\sqrt{2}-1}{7}$ and when the transmission parameter
$t$ satisfies
$$ t_-^2  \leq t^2 \leq  t_+^2,  $$
we have
\begin{equation}\label{condition1}
p^{2r^2}+ p^{2t^2} + 7 p^2 - 1 \leq 0,
\end{equation}
and the geometric quantum discord is given by
\begin{equation} D_{g}(\rho
_{AB})=\frac{\lambda _{1}+\lambda _{3}}{4} = \frac{1}{4}
p^2~\frac{p^{2t^2} + p^{-2t^2} + 6}{(1 + p^2 )^2} + \frac{1}{4}
p^{2(1-t^2)}\frac{(1-p^{2})(1-p^{2t^{2}})}{(1 + p^2 )^2}.
\end{equation}
For antisymmetric quasi-Bell states associated with $m = 1$ (mod 2),
the condition  $\lambda _{1} \leq \lambda _{2}$ is always satisfied
and in this case the geometric discord takes the sipmle form
\begin{equation}
D_{g}(\rho _{AB})=\frac{\lambda _{1}+\lambda _{3}}{4} =
\frac{p^{2r^2}(2 - p^{2t^2}- p^{2})}{4}\frac{1 - p^{2t^2}}{(1 -
p^{2})^2}.
\end{equation}
Here also, the geometric measure of quantum discord in the state
$\rho_{AE}$ is simply obtained from $D_{g}(\rho _{AB}$ modulo the
substitution $ r \longleftrightarrow s$.\\
 In the pure
bi-partitioning scheme $A\vert BE$, it is easy to check, using the
method presented in the previous subsection,  that the geometric
discord is related to the concurrence of the state $\rho_{A\vert
BE}$ (\ref{CABE}) as follows
\begin{equation}
D_{g}(\rho _{A\vert BE})=\frac{1}{2}C^{2}\big(\rho _{A\vert BE}\big)
\end{equation}
which can be written as
\begin{equation}
D_{g}(\rho _{A\vert BE}) =\frac{1}{2}\frac{(1-p)^{2}}{(1+p^{2}\cos
m\pi )^{2}}.
\end{equation}
\subsection{Monogamy of geometric discord}

To illustrate the above analysis, we shall consider the special case
where the decoherence of quasi-Bell cat states is simulated by the
action of a 50:50 beam splitter. We treat first  the evolution of
the geometric quantum discord for symmetric quasi-Bell cat states
$(m=0)$. In this case, using the results obtained in the previous
subsection, it is simply verified  that for $ 0 \leq p \leq
\frac{2\sqrt{2} - 1}{7}$
$$ D_g(\rho_{AB}) = D_g(\rho_{AE}) = \frac{p}{4} ~\frac{p^3 + 5p + 2}{(1+p^2)^2}$$
and for $ \frac{2\sqrt{2} - 1}{7} \leq p \leq 1 $
$$ D_g(\rho_{AB}) = D_g(\rho_{AE}) = \frac{1}{4}~ \frac{(1-p^2)^2}{(1+p^2)^2}.$$
We have also
$$ D_{g}(\rho _{A\vert BE}) =\frac{1}{2}\frac{(1-p)^{2}}{(1+p^{2}
)^{2}}$$ for $0 \leq p\leq 1$.  The behavior of the quantity
$$D_g(A,B,E) =     D_{g}(\rho _{A\vert BE}) - D_g(\rho_{AB}) - D_g(\rho_{AE}),$$
as function of the overlap $p$, is plotted in the figure 9.
\begin{center}
  \includegraphics[width=3in]{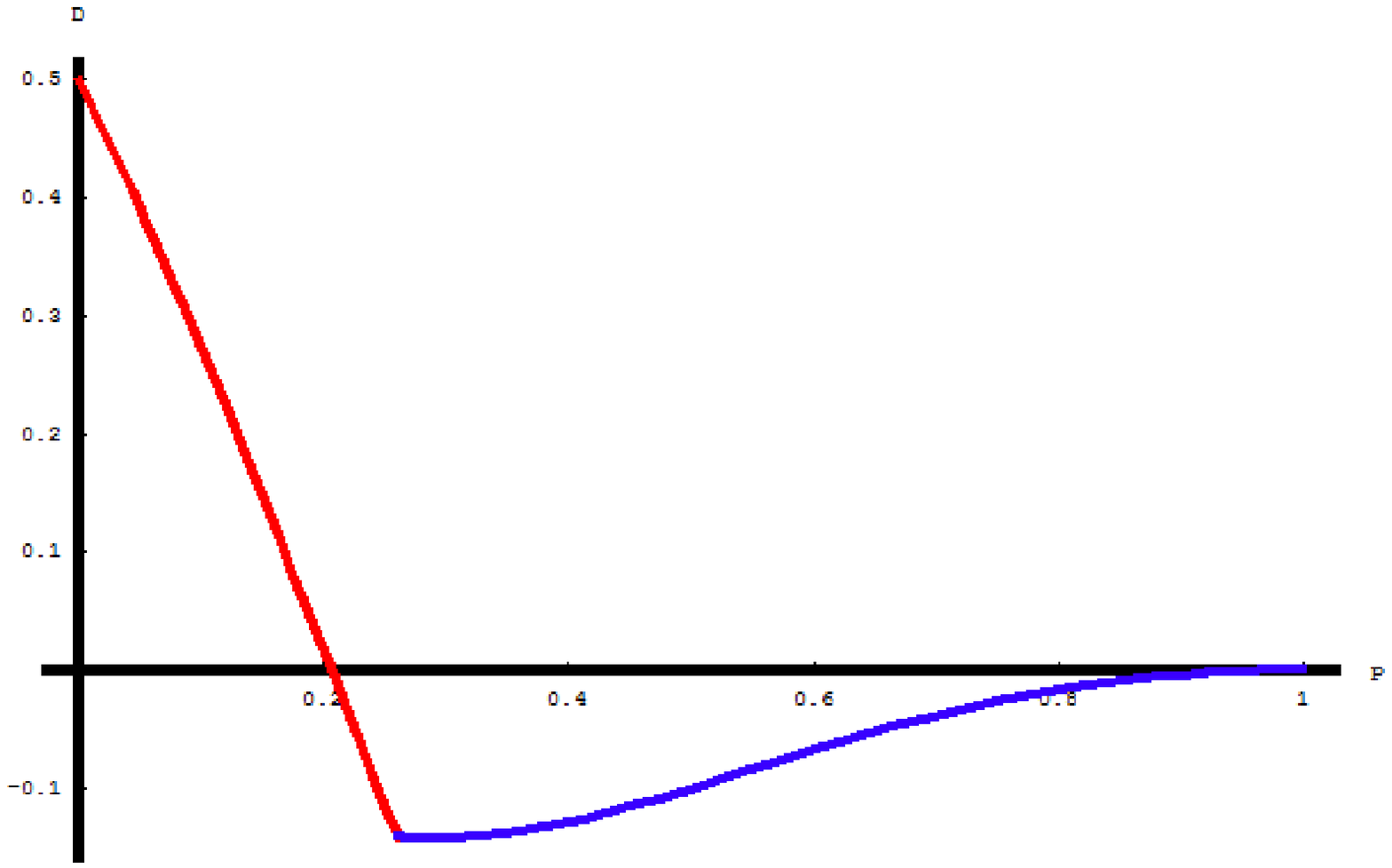}\\
{\bf Figure 9.}  {\sf  $D_g = D_g(A,B,E) $ versus the overlapping
$p$ for $t^2= \frac{1}{2}$ and   $m=0$.}
\end{center}
Clearly,  the geometric quantum discord is monogamous for quasi-Bell
cat states with $p$ such that $ 0 \leq p \leq
0.206783$, but does not follow the  monogamy property elsewhere. \\
For antisymmetric quasi-Bell states $(m = 1)$, we have
$$ D_{g}(\rho _{AB}) = D_g(\rho_{AE}) = \frac{1}{4}\frac{p^2 + 2p}{(1+p)^{2}}$$
and
$$ D_{g}(\rho _{A\vert BE}) =\frac{1}{2}\frac{1}{(1+p
)^{2}}.$$ In this case, the  quantity defined by $ D_g(A,B,E)$ is
positive for $ 0 \leq p \leq \sqrt{2} -1 $ and the geometric quantum
discord is monogamous. However, the monogamy is violated when $
\sqrt{2} -1 \leq p \leq 1$.

\section{Concluding remarks}

To summarize, we have studied the decoherence properties of
quasi-Bell cat states based on Glauber coherent states. The
decoherence effects are qualitatively modeled by the action of a
beam splitter. This effect is parameterized by a transmission
coefficient  $t$ to take into account the loss of the information
and subsequently the inevitable degradation of the quantum
correlations present in the initial system. We used a qubit mapping
to convert the continuous variables (even and odd Glauber coherent
states) to a discrete qubit setting. Through concurrence,
entanglement of formation, quantum discord and its geometrized
version, we characterized the quantum correlations between the two
modes of quasi-Bell cat states and the noisy channel. The explicit
analytic expressions of these measures were obtained. Finally, we
have investigated the distribution of entanglement of formation,
quantum discord and geometric discord between quasi-Bell cat states
and the environment. We have demonstrated that the quantum
correlations measured by squared concurrence satisfy the monogamy
relation. However, when the correlations are measured by means of
based-entropy measure like entanglement of formation and quantum
discord or  distance-based measure as the geometric quantum discord,
the monogamy is satisfied  in some particular cases depending on the
strength of the coupling to the environment which is characterized
by the parameter $t$ and the overlapping $p$ of the Glauber coherent
associated with  the quasi-Bell cat states under consideration.
Especially, for each of above mentioned measures, we determined the
critical values of transmission parameter $t$ and overlap $p$ under
or below which the monogamy relation is
satisfied or violated.\\
The analysis presented here can be extended in many ways. For
instance, it is readily generalizable to quasi-Bell cat states based
on spin coherent states as well as coherent states associated with
other Lie algebras.  It will be also an important issue to extend
these results to others mechanisms inducing decoherence effects.
Further thought in this direction might be worthwhile.

\end{document}